# APPROACHES TO AUTOMATED MORPHOLOGICAL CLASSIFICATION OF GALAXIES

Avi Naim

Institute of Astronomy, Madingley Rd.,

Cambridge, CB3 0HA, U.K.

e-mail : hn@mail.ast.cam.ac.uk

**Abstract** – There is an obvious need for automated classification of galaxies, as the number of observed galaxies increases very fast. We examine several approaches to this problem, utilising *Artificial Neural Networks* (ANNs). We quote results from a recent study which show that ANNs can classsify galaxies morphologically as well as humans can.

## 1. INTRODUCTION

Morphological classification of galaxies is usually done by visual inspection of photographic plates. This is by no means an easy task, requiring skill and experience. It took years to compile catalogues of galaxies containing of the order of $10^4$ entries. However, the number of galaxy images available today or in the near future is two orders of magnitude larger. Clearly, such numbers of galaxies cannot be classified by humans. There is an obvious need for automated methods that will put the knowledge and experience of the human experts to use and produce very large samples of automatically classified galaxies.

There are three issues which need to be settled at the outset :

1. On which features do we base the classification procedure ?

2. Which automated classifier do we want to use ?

3. What is the criterion for measuring our success ?

While the latter two questions are technical, the first is much deeper. When setting out to construct an automated classification procedure we first ought to define our ultimate goal : Do we want to replicate existing classification schemes on many unclassified galaxies, or do we wish to come up with a new scheme ?

Extending existing classification schemes to many unclassified galaxies calls for an approach known as *Supervised Learning* : An automatic procedure that "learns" from examples and is able to use the "knowledge" it acquires on large quantities of galaxies it has never been presented with before. Alternatively, there is the "unsupervised" approach, whereby we want to perform some "un-prejudiced" classification. Ideally, such a classifier would be presented with raw images of galaxies, and asked to group them together on the basis of some pre-defined similarity criterion. If such a classifier is then presented with thousands of examples it may be able to come up with a whole new classification system, i.e. create its own version of the Hubble Sequence.

Our preferred type of classifier is Artificial Neural Netrworks (ANN), which proved to be a promising tool in a pilot study (Storrie-Lombardi *et al.* 1992). This is by no means the only possible choice, but there are certain attractive statistical features to the ANN approach, which are discussed elsewhere (see Lahav in this volume). When constructing an ANN one has to specify its architecture in full, describing the number of input parameters, the number of output nodes and the number of nodes in the hidden layer. The number of input parameters depends on how we choose to describe a galaxy. The number of output nodes depends on the desired format of classification. Both of these aspects are discussed below, in § 2 and § 3, respectively. We bring some examples and recent results in § 4 and summarise briefly in § 5.

## 2. CHOICES OF INPUT PARAMETERS

The number of input parameters ought to be restricted for practical as well as theoretical reasons. From a practical point of view, too many input parametes increase the complexity of the network. The more nodes there are, the more connections there are between nodes. Each such connection has a strength, called "weight", the weights in a given ANN being its "free parameters". Since the ANN tries to minimise its classification error in the multi-dimensional space spanned by all of its weights, the more weights we have the higher the likelihood of the network to get stuck in a local minimum of its error and not reach the global minimum. One may view each galaxy as a class in its own, but then the whole concept of classifying into a limited number of categories breaks down. If we want to end up with only a few classes, we need to present the ANN with those parameters *which tell classes* *apart*. When a human expert looks at the image of a galaxy on a photographic plate she (or he) does not take each picture element separately into account, but probably looks for specific correlations (e.g. arms) of picture elements or some overall variability (e.g. the concentration of light).

For the purpose of *morphological* classification we therefore ought to find means of compressing the information held in thousands of picture elements to a small number of parameters (ideally of order 10). If the ANN is to be supervised these parameters ought to correspond to well known indicators of type (e.g. Bulge–to–Disk ratio, development and tightness of the arms, etc.). We refer to this kind of information as "subjective". On the other hand, if the ANN is to be unsupervised, this type of information will bias it towards our preconceptions of galaxy morphology. For this purpose it may be better to find a more "objective" compression technique, which ends up conveying the same information as the raw image - but with a lot less parameters. There are many compression techniques, but not all are useful to this end. If humans recognise patterns (not only of galaxies) by inspecting correlations between different parts of them, then "smart" compression algorithms which treat a picture merely as a sequence of bits are of little use here. Techniques such as Principal Component Analysis (PCA), Fourier transforms or Wavelet transforms are better suited for this problem.

Understanding galaxies, however, is not just about morphologies. As more and more data become available about the dynamics and chemical composition of galaxies it becomes more feasible to utilise this information in classifying galaxies. Colour indices, for example, were very useful to the ANN which classified galaxies from the ESO catalogue (Lauberts & Valentijn, 1989), as shown in Storrie-Lombardi *et al.* (1992). The supervised ANNs may well benefit from the inclusion of such data, but their importance will probably come into full view when used in unsupervised ANNs. There we may discover aspects of the physics of galaxies which were hidden hitherto, in the spirit of the "fundamental plane" (e.g. Bender, Burstein & Faber 1992). The major problem here is that most galaxies will have only some of the information available, and the ANN will have to deal with missing data in its input nodes.

## 3. CHOOSING THE NUMBER OF OUTPUT NODES

The simplest ANNs utilise a single output node, which takes on a T-type value for each galaxy, along a given scale (e.g. from -5 to 10 in the Revised Hubble Sequence, de Vaucouleurs *et al.* 1991). Using a single output node corresponds to assuming that galaxies indeed form a sequence in the space defined by the input parameters, and so the ANN is never going to have to "make its mind up" between classifying into radically different types (e.g. E+

or Sbc), forcing the output node to take on some intermediate value. This may be true for purely morphological classifications. However, in the more general context different parameters may result in conflicting tendencies in the ANN. A single output ANN will be unable to cope with such a case, since it adopts one value only for each galaxy. The way around this difficulty is to use several output nodes, each of them representing one output class. In this case the ANN can be forced to approximate in its output nodes the Bayesian *a posteriori* probabilities of the galaxy to belong to their respective classes (although there is an ad-hoc binning into several classes, to begin with). Cases of indeterminate classification will then come out naturally as large probabilities in two different nodes, which do not necessarily next to each other.

Another kind of multiple–output ANNs is predicting not only the T-type, but some other parameters as well (e.g. Luminosity class). In this kind of configuration the idea is to get the ANN to tell us several different things about the object it is presented with, all in one go. For example, we may require both the T-type and the Luminosity class of a galaxy.

Yet another kind of ANN which uses multiple–outputs is known by the name "encoder". This is essentially an unsupervised method, although it uses the mechanism of a supervised ANN. The ANN is presented with the input parameters and is required to reconstruct them at the output layer, i.e., there are as many outputs as there are inputs. Typically, the number of hidden nodes used is much smaller, so the ANN effectively compresses the data in the inputs into some representation in the hidden layer. The values of the hidden nodes can then be plotted against each other to see whether there is a segregation of the data points into different groups.

## 4. AN EXAMPLE : MORPHOLOGICAL CLASSIFICATION OF APM GALAXIES

The example described below is given in full detail elsewhere (Naim *et al.* in preparation). Here we only give a brief description designed to serve as an example of some of the considerations discussed above. For this project we used a diameter–limited ($D > 1.2\ arcmin$) sample of 835 galaxies taken from the APM Equatorial Catalogue of Galaxies (Raychaudhury *et el.* in preparation). The plates were obtained with the 48" UK Schmidt telescope at Siding Spring, Australia, and digitised by the Automated Plate Measuring (APM) machine in Cambridge. We first sent laser prints of the images to six experts (R. Buta, H. Corwin, G. de Vaucouleurs, A. Dressler, J. Huchra and S. van den Bergh) for classification. The analysis of their work was reported elsewhere (Naim *et al.* 1994) and supplied us with a basic classified sample on which to train the ANN. We used both a single–output ANN for predicting the T-type and a 16–output ANN with probabilistic outputs, as discussed above. We chose to use 16 outputs in order to span the full range of types from −5 to 10 on the Revised Hubble System.

The images were all reduced using dedicated automatic software written for this purpose and then sampled on 30 ellipses, all with the same ellipticity and position angle as the entire image. This sampling method provided us with a standard set of measurements for all galaxies, regardless of angular size, tilt and position angle. This standard set contained roughly 6000 points for each galaxy. We then looked at ways of extracting few significant features from this set. In figure 1 we show two galaxies from our sample, one being an early type galaxy and one being a spiral. The plots of the sampled ellipses of these galaxies are shown at the top of figure 2 (innermost ellipse is bottom, outermost is top), and the bottom two plots in figure 2 show their light profiles (average intensity along a given ellipse vs. ellipse number). It is plain to see in these plots which is the early type galaxy and which the spiral : The spiral has a smaller bulge (compared to the overexposed central region in the light profile of the early–type galaxy) and the outer half of its light profile is much flatter. The ellipses themselves show very little correlated, long–range structure in the early–type galaxy, whereas the arms of the spiral clearly stand out. Note, however, that none of the statements just made can result from observing the image one pixel at a time. We therefore prepared dedicated software to extract features, motivated by the criteria the human experts use.

We ended up with 24 parameters which were then compressed to 13 using PCA. We ran many ANNs and took the average of all runs, in order to exclude accidental results. we found that the ANN's rms dispersion relative to the mean types worked out from the classifications of the individual experts was 1.8 types (on the 16 type scale of the Revised Hubble System). For comparison, the overall rms dispersion between the experts themselves was also 1.8 types. This implies that our choice of input parameters for the ANN indeed gave a good description of those morphological features that are important for classification. One may claim, however, that this success if due *in full* to the choice of input parameters and that the choice of classifier does not play a part in its success. It was clear from our results that the number of hidden nodes used for given numbers of inputs and outputs was of minor importance, but in order to check whether the non-linearity of the ANN played any significant role we also ran a fully linear ANN with no hidden nodes. There the rms dispersion was found to be 2.2 types, and this meant that the choice of a non-linear classifier is of great importance, and in general one may not dismiss the classifying method as an unimportant element in the process.

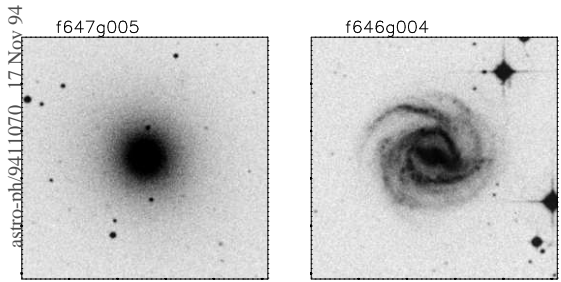

Figure 1: Images of two galaxies from our sample

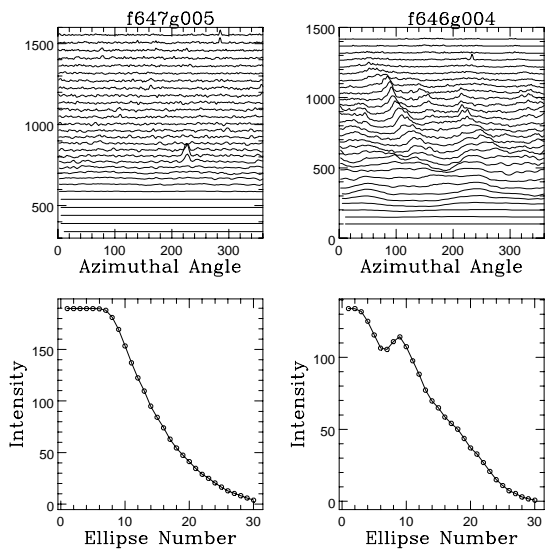

Figure 2: The sampled ellipses (top) and light profiles (bottom) of the two galaxies whose images appear in the previous figure

## 5. SUMMARY AND FUTURE WORK

We have discussed the general motivation for using an automated galaxy classifier. In the context of ANNs we have explained the importance of tailoring the parameters presented to the ANN to the problem in question, and discussed various choices of the number of output nodes. An example of a supervised ANN motivated by morphological features was given. In the near future we intend to use that same sample of 830 galaxies in order to apply other supervised methods (e.g., using physical parameters as well as the morphological features) and unsupervised methods to the problem of galaxy classification.

### Acknowledgements

I thank the APM group at RGO Cambridge for scanning support, and especially my collaborators O. Lahav, L. Sodré Jr. and M. C. Storrie-Lombardi.